\documentstyle[psfig]{caosp}

\begin{document}

\htitle{Doppler-Zeeman mapping ...}
\hauthor{V.L. Khokhlova et al.}
\title{Doppler-Zeeman mapping of magnetic CP stars: 
        The case of the CP star HD 215441 }
\author{ V.L.Khokhlova   \inst{1} \and
         D.V.Vasilchenko \inst{2} \and
         V.V.Stepanov    \inst{2} \and
         V.V.Tsymbal     \inst{3} }
\institute { 
    Institute of Astronomy, Russian Academy of Sciences,
       Pyatnitskaya ul. 48, Moscow, 109017 Russia. \and
    Department of Computational Mathematics and Cybernetics, 
       Moscow State University, Moscow, 119899 Russia. \and
    Simferopol State University, Department of Physics,
       Simferopol, Ucraine. }

\pubyear{1998} \volume{28} \firstpage{999}
\offprints {V.L. Khokhlova, $e-mail$: vlk@khokh.msk.ru }
\date{}
\maketitle

The recently developed method of Doppler-Zeeman mapping of magnetic CP stars
(Vasilchenko et al.,1996, hereafter referenced as V.S.K., and also the
contribution to this volume) confirmed its efficiency
and robustness not only for model tests but also in the case of a real star.

The method  was used for mapping Babcock's star HD 215441, wich has the 
strongest known magnetic field and resolved Zeeman components in the lines of 
its spectra (Khokhlova et al., 1997, hereafter referenced as K.V.S).

We used the spectra of the CP star HD 215441 kindly given to us by Prof.
Landstreet and which he had obtained with the coud\'e-spectrograph of the 3.5m
Canada-France-Hawaii telescope and a Reticon detector. The spectral resolution 
was 0.1 \AA\, and the signal-to-noise ratio $S/N \ge 200$. We did not have
polarization data at our disposal, but completely resolved
$\pi$ and $\sigma$ components are practically equivalent to measured Stokes
V parameter (except for the sign of the polarization and hence the
lack of possibility to determine the sign of the magnetic-dipole vector).
We assumed the signs of poles obtained in previous studies of this star
(Landstreet et al., 1989).

Abundance maps for Si, Ti, Cr and Fe obtained from different lines of each
element coincide very well. The magnetic field configurations obtained from 8
lines of four different elements were identical. They have been reliably
determined for the star's surface facing the observer. The maps are
shown in Fig.~1.

The magnetic field configuration is well fitted by a shifted-dipole model
with dipole parameters $H_D=(21.3 \pm 1.9)$ kG and shift 
$a=(0.12 \pm 0.024) R^*$.
The positive polar field $H_{p^+} = +(39.0 \pm 2.0)$ kG and the negative 
$H_{p^-}= -(44.0 \pm 9.0)$ kG. The local magnetic vectors were calculated for 
each point of the stellar surface.
The position of maximum surface magnetic field $H_S$ on the visible hemisphere
is determined to be $L=(5^\circ.6 \pm 6^\circ.8)$, 
$\varphi=(63^\circ \pm 7^\circ.8)$ and the strength of the
visible magnetic pole is $max (H_S)=(42.3 \pm 2.9)$ kG.

A comparison of magnetic and chemical maps definitely shows that at the visible
magnetic pole most elements are strongly (Si, Cr) or slightly (Ti)
underabundant but Fe is strongly overabundant. A large-scale ring structure
around the magnetic pole is clearly seen on abundance maps (Fig. 1).
The complete ring with strong Si overabundance is situated at low magnetic
latitudes but not exactly on the magnetic equator. A strongly overabundant
spot of Ti and of Cr is located on the background of a less overabundant ring,
which is at the same place as the strong Si (and faint Fe) ring. Its location
is particular, being not only near the magnetic equator, but also
on the rotational equator.

The possibility of such a detailed and reliable comparison between magnetic 
field structure and abundance maps for magnetic CP stars is demonstrated for the
first time in this study. The mapping became possible with a modest computer
facility (PC Pentium) thanks to the method we used: analytic approximation of 
local line profiles and series expansion of the second order of multipole 
magnetic field geometry. The magnetic field of this star could be well
approximated by a shifted dipole as well as by centered dipole and quadrupole.

We had to assume that the atmospheric model parameters $T_{\rm eff}$ and
$\log g$ are constant over the surface of the star but this may not be true in 
reality. Due to evidently strong chemical inhomogeneities seen in Fig. 1, one 
may expect different metallicities at different places. But it is well known,
that Kurucz's models with different metallicities may provide big differences
in computed line intensities. For our case it is demonstrated by Fig. 4 of
K.V.S..

We cannot tell as yet, which element's abnormal abundance is most
efficient to produce such effects, because Kurucz's metallicity
is a scaled Solar abundance. One may only guess that C, N, O, Si, as most 
abundant elements, may play a dominant role as electron donors at the 
temperature of this star ($T_{\rm eff}=16000$) K, and also Fe, which is abundant
and has many lines to contribute to the backwarming effect.

In general, the strict approach to solve the inverse problem of Doppler-Zeeman
mapping requires iterations to be built as follows:

At each step of minimization of residual Functional, one uses the values of
local abundances, local atmosphere models and local magnetic field vectors
all obtained at the preceding step, and then performs the integration
of the transfer equation to compute local Stokes profiles and local
surface brightness, integrates them over the visible surface to calculate
the "observed" profile and the residual Functional for the next step of
minimization.
Obtaining the new set of local abundances, one should make computations
of new local atmosphere models.
Now the ``loop'' is closed and a new iteration may be started.

Keeping in mind that a dense coordinate grid on the stellar surface is desirable
and that local abundances of more than one element are needed to be determined
to make corrections to local atmosphere models at each iteration, the amount
of time required for the strict solution by the above scheme is huge and
unrealistic with the methods known and the available computers at the
present time.

\begin{figure}[hbtp]
\centerline{
\psfig{figure=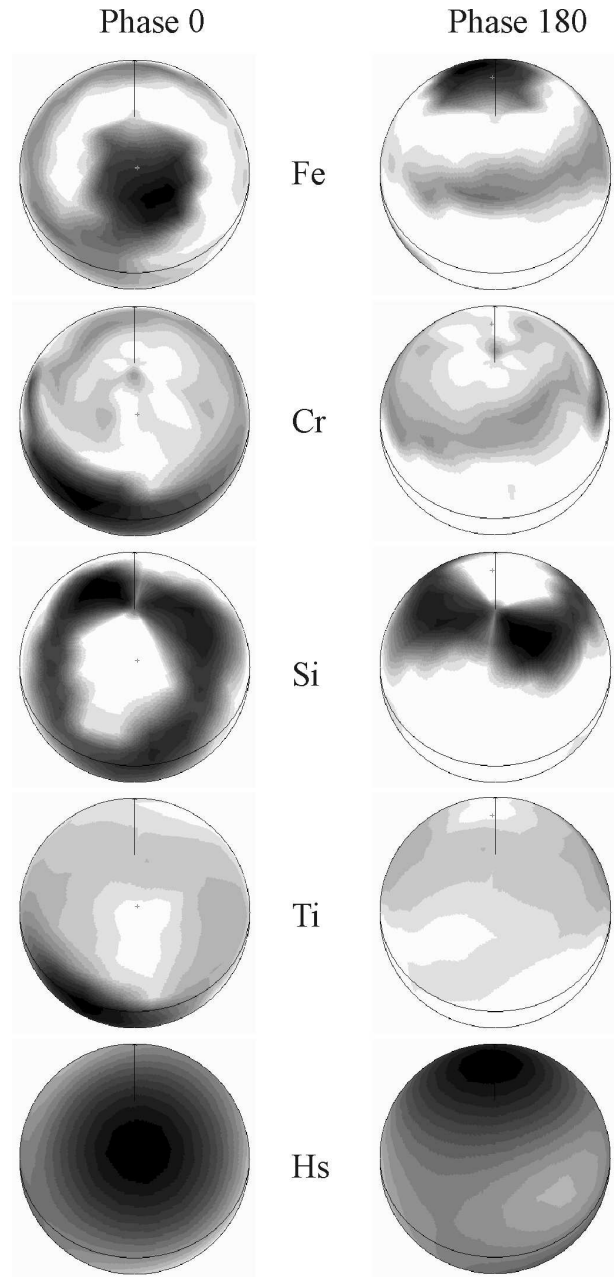,width=8cm}}
\caption{Chemical maps for Fe, Cr, Si and Ti and map of
distribution of the local surface field $H_S$ on HD 215441 at phase
$0^\circ$ (left column) and $180^\circ$ (right column)}
\label{fp}
\end {figure}

The way we compute local Stokes profiles as a corrected analytic solution
of transfer equations (as is described in detail in V.S.K) opens the
possibility to solve the problem in a realistic time without serious
loss of precision. It seems that one may not blame this method of
local profiles computing for
having not enough precision, until the problem of determining correct
local atmosphere models is solved. It is impossible to guarantee better
local profiles by strict numerical integration of the transfer equation
only, without solving this problem.
We have also to mention the problem of vertical stratification
of abundances in CP stars, predicted by theoreticians (Babel \& Mishaud, 1991)
and investigated by observers (Babel, 1994; Romanyuk et al., 1992, and other 
attempts). If this stratification exists, further complications appear.

Returning to Fig. 1, one may see that Fe is overabundant near the magnetic
pole. If a ``metallicity'' effect is determined by iron abundance, the
iron spot must be hotter and there must be a bright spot there. Indeed,
it is known that the maximum of the almost sinusoidal lightcurve of HD 215441
occurs at zero phase.

One may expect that the local intensity of Ti\,{\sc ii} and Cr\,{\sc ii}
lines should decrease in a hot spot due to the second ionisation of these ions, 
and it is really the case, as Fig. 1 shows. But one cannot tell, whether this
is the only reason, or whether real underabundances of Ti and Cr also exist.

For Si\,{\sc iii} lines the temperature change has the reverse sign, their 
intensity should increase with increasing metallicity (increasing temperature
in line-forming layers). But they decrease! This may be interpreted only as a
real underabundance of Si near the magnetic pole.

A more detailed study of HD 215441 from spectra covering a wider wavelength
range, involving more lines and other elements and also including polarization
spectra, is urgently needed and seems to be very promising.

\acknowledgements
We thank Prof. J. Landstreet for giving us his
reticon spectra of HD 215441.
This work has been supported in part by the Soros International Science
Foundation and the Government of the Russian Federation (grants nos. N2L000
and N2L300) and by the program ``Astronomy'' (project no 3-292).

\end{document}